\documentclass[12pt,preprint,preprintnumbers,amsmath,amssymb,prb]{revtex4}
\usepackage{graphicx}

\begin{document}

\title{Aharonov-Bohm effect in the tunnelling of a quantum rotor \\in a linear Paul trap}

\author{Atsushi~Noguchi$^{1}$}
\email[]{noguchi@qe.ee.es.osaka-u.ac.jp}
\author{Yutaka~Shikano$^{2,3}$}
\author{Kenji~Toyoda$^{1}$}
\author{Shinji~Urabe$^{1}$}
\affiliation{%
$^{1}$Graduate School of Engineering Science, Osaka University, 1-3 Machikaneyama, Toyonaka, Osaka 560-8531, Japan.}
\affiliation{%
$^{2}$Research Center of Integrative Molecular Systems (CIMoS),Institute for Molecular Science, 38 Nishigo-Naka, Myodaiji, Okazaki 444-8585, Japan.}
\affiliation{%
$^{3}$Institute for Quantum Studies, Chapman University,1 University Dr, Orange, CA 92866, United States.}

\date{\today}

\begin{abstract}
Quantum tunnelling is a common fundamental quantum-mechanical phenomenon that originates from the wave-like characteristics of quantum particles. 
Although the quantum-tunnelling effect was first observed 85 years ago, some questions regarding the dynamics of quantum tunnelling remain unresolved. 
Here, we realise a quantum-tunnelling system using two-dimensional ionic structures in a linear Paul trap. 
We demonstrate that the charged particles in this quantum-tunnelling system are coupled to the vector potential of a magnetic field throughout the entire process, even during quantum tunnelling, as indicated by the manifestation of the Aharonov-Bohm effect in this system.
The tunnelling rate of the structures periodically depends on the strength of the magnetic field, whose period is the same as the magnetic-flux quantum $\phi _0$ through the rotor [($0.99 \pm 0.07 )\times \phi _0$]. 
\end{abstract}

\maketitle

The quantum dynamical phenomenon in which a particle tunnels through
a barrier that it classically could not penetrate\cite{Tunnel} was first
investigated in the 1920s to explain alpha decay\cite{Gamow, Condon}. 
There are also many other natural phenomena ranging from cosmological scales\cite{Vilenkin} to microscopic scales\cite{Kuki}, such as
chemical reactions\cite{Hanggi}, that arise from the quantum-tunnelling effect; 
quantum-tunnelling effects in artificial materials have been investigated using diodes (tunnel diodes\cite{Esaki}) and superconductors (Josephson
junctions\cite{Josephson2, Giaever2, Giaever, John}) and have been achieved in quantum simulators\cite{Upcroft, Roos}. 
Determining the dynamics of quantum tunnelling is a difficult task because the
momentum during quantum tunnelling might be pure imaginary\cite{Chiao},
and it is the tunnelling probability that is usually calculated
instead\cite{Landauer, Davie}. 
However, the dynamics of tunnelling particles can be investigated if one can construct a highly coherent tunnelling system.

Trapped ions form a highly controllable, isolated quantum system in which both the internal and motional degrees of freedom have long coherence times\cite{James}. 
Using one-dimensional ion chains trapped in a linear Paul trap, the oscillation can be cooled to near the ground states, and phonon-manipulation can be achieved through laser irradiation\cite{Wineland}.
An important feature of a trapped-ion system is the Coulomb interaction between ions, which generates an effective interaction between ions for quantum information processing\cite{James}.
By virtue of these characteristics, a trapped-ion system is one of the more promising candidates for quantum information processing\cite{James}.
However, the Coulomb interaction also provides several types of spatial arrangements of the trapped ions, i.e., Wigner crystals, including two-dimensional structures\cite{Fishman}.
So far, the phase transitions\cite{Morigi, Plenio} and dynamics\cite{Partner} of these Wigner crystals have been investigated both theoretically and experimentally.
However, these experiments have not been performed in the quantum regime\cite{Partner, Wunderlich} because the motions of such two-dimensional structures have never been cooled to near the ground states.

The Aharonov-Bohm (AB) effect\cite{Aharonov}, in which a charged particle is affected by a magnetic field even as it travels  through a region in which the magnetic field is zero, has been experimentally confirmed using an electron in a superconducting ring\cite{Tonomura, Tonomura2} and other materials\cite{Bachtold, Morinaga, Tarucha, Peng}. 
Even in a non-localised magnetic field, the wave function of a charged particle obtains a phase that is proportional to the area of a closed loop\cite{Aharonov2, Mooij}. This effect is also referred to as the AB effect. 
A charged quantum-tunnelling particles should be affected by the vector potential of a magnetic field even during quantum tunnelling.
Thus, the AB effect should occur for tunnelling particles and is an example of the dynamics of tunnelling particles.

In this study, we have realised a quantum-tunnelling rotor, that consists of three ions and has a two-dimensional structure that is cooled to near the motional ground state in a linear Paul trap.
This rotor has two stable orientations and can transition from one to another though quantum tunnelling.
Because of the controllability and simplicity of the trapped-ion system, this quantum-tunnelling rotor may offer a novel approach to understanding quantum-tunnelling dynamics, and we were successful in observing the AB effect of tunnelling particles using this system.
Several recent technical advancements regarding the manipulation of
ions, such as gate operations and coherent motional controls\cite{James, Wineland}, offer further means of controlling the quantum-tunnelling rotor.

\section*{Results}
\subsection*{The quantum-tunnelling rotor}
A linear Paul trap, whose normal-mode frequencies are comparable along two axes (x,z) and large along the remaining axis (y), can be used to trap a single triangular structure of three $^{40}\mathrm{Ca}^+$ ions (Fig. 1a)\cite{Wineland, Fishman}. 
The symmetry of the potential defines two stable orientations for the triangular structure, which are referred to here as `up' and `down', and the
motional modes of these structures are more complex than those of a one-dimensional Coulomb chain\cite{Morigi}. 
Of the four lowest energy modes represented in Figure 1b, the mode indicated by the solid curve represents a `rotational mode' in which the ions rotate around the centre of the triangular structure. 
Here, we demonstrate the cooling of the rotational mode, which is one of the two-dimensional modes, to near the ground state.
Theoretically, the rotational mode has interesting characteristics: its normal-mode frequencies become zero at critical points, which corresponds to $\omega _x=2\pi\times 1.119$ $\mathrm{MHz}$ in Figure 1b. 
Under such conditions, the rotational mode becomes massless, and singular phenomena that are similar to other degrees of freedom of two-dimensional structures in a linear Paul trap are expected\cite{Plenio}.

Near the critical point of the rotational mode (Fig. 1b), the wave
functions of each ion in the rotational direction are elongated, and
quantum tunnelling can occur between two degenerate triangular structures\cite{Gunther}. 
Because the other motional modes have much larger spring constants
by virtue of the strong Coulomb interaction, it is reasonable to assume
that these modes are unaffected by the quantum tunnelling. These
triangular structures become a quantum rigid rotor, which is a quantum system
that has a rotational degree of freedom but maintains its structure. 
In this system, the rotation of the quantum rigid rotor
is driven by the quantum tunnelling.
We refer to this system as a `quantum tunnelling rotor' (Fig. 1).
Ions also have internal degrees of freedom, i.e., spins, and the tunnelling can depend on the spin states of the ions of the quantum-tunnelling rotor\cite{Gunther}.
Using this system, we can investigate and manipulate the tunnelling system itself by controlling the trap parameters and the spin state.
Figures 1c and 1d illustrate the effective potential of the quantum-tunnelling rotor in the linear Paul trap for the actual experimental parameters used.
Each bound state in the periodical multi-well potential corresponds
to a single stable orientation of the quantum-tunnelling rotor. In
addition, the overlap of the up and down wave functions of the ions
observed in Figure 1d indicates that quantum tunnelling is possible. The
orientations of the atomic triangles in this condition are not
stable under laser-cooled temperatures.
However, under intermediate radial-confinement conditions, corresponding a normal frequency of $\omega _x=2\pi\times 1.523$ $\mathrm{MHz}$in Figure 1b, both degenerate triangular structures are stable even for `hot' Doppler-cooled temperatures. 
Thus, projective measurements of the triangle orientations can be performed.

\subsection*{Observation of quantum tunnelling}
We experimentally observed the quantum tunnelling of the rotor using a magnetic field of a certain magnitude. 
The three ions were trapped in a linear
Paul trap with normal-mode frequencies of $\{ \omega _x,\omega
_y,\omega _z\} =2\pi\times \{ 1.523, 1.961, 1.119\}$ MHz (further
details of the experimental system can be found in Ref. \cite{noguchi, haze}),
and the orientation of the triangular structure could be set using intermediate-confinement conditions. 
Figure 2a shows the time sequence of the experiment, which began with the measurement of the
initial state of the triangular structure using a charge-coupled
device (CCD) camera. 
For quantum tunnelling to be observed, it was necessary to cool the quantum-tunnelling rotor to the rotational ground state. 
The rotational mode had the lowest heating rate among the two-dimensional motional modes (see the Methods section) and could be cooled to near the motional ground state (a mean phonon number of $n_{rt}=0.088\pm0.007$) using the sideband cooling technique.
After cooling, the trap radio frequency (RF) was adiabatically reduced to create the quantum-tunnelling rotor.
At that time, the normal-mode frequency of the rotational mode changed to $2\pi\times 0.18\mathrm{kHz}$ from $2\pi\times 750\mathrm{kHz}$ (Fig. 1b).
There was a small amount of heating of the quantum-tunnelling rotor caused by noise, micromotion, and diabaticity when the trap potential was
changed. Under tunnelling conditions, we assumed the mean phonon number
to be approximately $n_{rt}\sim 4$ (see the Methods section).
After a time $\tau$, the trap RF was increased, and we again measured the orientation of the triangular structures. 
Near the critical point, the quantum tunnelling rotor in the rotational ground state oscillated between the two degenerate orientations because of the quantum tunnelling (Fig. 2b).

\subsection*{Magnetic-field dependency of quantum tunnelling}
The AB effect must occur in the quantum-tunnelling rotor in a magnetic field because the dynamics of the quantum-tunnelling rotor are governed
by the quantum interference of two-directional rotations; see the Methods section for details.
Thus, the tunnelling of the quantum-tunnelling rotor can be controlled by adjusting the strength of the magnetic field.
The dependency of the transition probability on the number of magnetic-flux quanta was investigated by fixing the waiting time to $50$ $\mathrm{ms}$ and adiabatically varying the magnetic field, with the field held constant during tunnelling (Fig. 3). 
The detail of the experimental setup regarding the magnetic fields are presented in the Methods section.
The oscillation period estimated from a sinusoidal fit is $(0.99\pm 0.07)\phi_0$, for the single magnetic-flux quantum $\phi _0 =h/e$, where $h$ is the Planck constant and $e$ is the elementary charge.
This fitting result agrees with the theoretically predicted oscillation period obtained from Eq. (6), which is presented in the Methods section.

\section*{Discussion}
Figure 2b presents the experimental results for the tunnelling dynamics of the rotor.
These dynamics can be divided into quantum tunnelling and classical rotation.
The ions are not completely in the ground state; ions in the rotational excited states rotate without tunnelling because the energies of these states are 
greater than the potential barrier, whose height is $h\times 250\mathrm{Hz}$ (Fig. 1d) where $h$ is the Planck constant.
The transition probabilities of the tunneling consist of two components.
One component represents the tunnelling between the up and down states and can be described by a sinusoidal curve, and the other represents the rotation  of the excited states and can be described by an exponential decay.
A fit of the experimental data to a sum of
these two curves yields a tunnelling rate of $7.6\pm 0.3$ $\mathrm{Hz}$ and a decay time constant of $5.4 \pm 0.3$ $\mathrm{s^{-1}}$, i.e., the mean rates of the classical rotation.
The amplitude of the oscillation, which was determined to be 0.19 based on the thermal
distribution of the phonon number, is consistent with the
population of the ground state. 
Furthermore, our tunnelling rates for different radial confinements (Fig. 2c) exhibit good agreement with the numerical solution of the Schr\"{o}dinger equation for the potential illustrated in Figure 1d. 
The oscillation in the transition probability can therefore be attributed to quantum tunnelling.

Figure 3 shows the strength of the magnetic-field dependency of the tunnelling rate.
This figure demonstrates that the ions couple to the vector potential even under quantum-tunnelling conditions; i.e., the system exhibits the AB effect. 
The amplitude of the oscillation is also consistent with the quantum tunnelling result (Fig. 2b). 
Note that the AB effect does not occur in classical rotations of the excited states because of the incoherency.
The quantum-tunnelling rotor may be affected by the Lorentz force during quantum tunnelling.
The Lorentz force can be calculated from a velocity and a magnetic field.
The expectation value of the particle position can be evaluated using Ehrenfest's theorem;
however, the velocity of a tunnelling particle is expected to be purely imaginary because $E-U$ is negative (where $E$ is the kinetic energy and $U$ is the potential energy). 
One approach to circumventing this difficulty is to take the absolute value $|E-U|$ to be the tunnelling kinetic energy.
The validity of this approach is supported by quantum-clock conditions\cite{Landauer}, the Ehrenfest theorem\cite{CL}, and an analogous experimental demonstration using a single photon\cite{Chiao}.
Using this approach, we estimate the effective maximal rotation speed $v$ of the rotor to be $\sqrt{2\vert U_0 - E \vert /M}$\cite{Landauer, Chiao, CL}. 
Here, $U_0 =h\times 270$ $\mathrm{Hz}$ is the maximal potential energy (Fig. 1d), $E=h\times 90$ $\mathrm{Hz}$ is the energy of the rotational ground state, and $M=3\cdot 40\cdot 1.67\times 10^{-27}$ $\mathrm{kg}$ is the mass of the quantum-tunnelling rotor. 
This speed yields a maximal strength of the Lorentz force of $evB \sim 10^{-26}$ $\mathrm{N}$ for a magnetic field of $B\sim 5$ $\mathrm{G}$.
An alternative approach is to calculate the effective mean velocity $v_m$ from the tunnelling rate. In this case, $v_m = Jr_0\pi /3$,
where $J=7.4$ $\mathrm{Hz}$ is the tunnelling rate and $r_0=3.42$ $\mathrm{\mu m}$ is the radius of rotation. This velocity yields a mean strength of the Lorentz force of $ev_m B\sim 10^{-27}$ $\mathrm{N}$ for the same magnetic field.
According to the results of both approaches, the Lorentz force is quite small and can be ignored in our experiment because the Lorentz-force-induced change in the radius of the closed loop would be on the order of femtometres.

\section*{Methods}
\subsection*{Quantum-tunnelling rotor in a magnetic field}
The AB effect occurs in the quantum-tunnelling rotor, and the wave functions of the ions gain a phase that it proportional to the homogeneous magnetic field and the area of the closed loop drawn by the path of each ion of the quantum-tunnelling rotor\cite{Aharonov2, Schulman}. 
Because of the strong radial confinement of each ion, this closed loop corresponds to the potential valley shown in Figure 1c.

The wave function of the quantum tunnelling rotor can be expressed as\cite{Oshikawa}
\begin{equation}
\psi _{up}(t+\Delta t)=\psi _{up}(t)+i(J_r +J_l) \Delta t\psi _{down}(t),
\end{equation}
where $\psi _{\mathrm{up,down}}(t)$ are the wave functions of the
two quantum-tunnelling rotor orientations, $\Delta t$ is a short
time-interval, and $J_{r}$ ($J_{l}$) is the complex amplitude of the
tunnelling for clockwise (anti-clockwise) rotation. Because the AB
effect introduces a phase difference between the clockwise and
anti-clockwise rotations (Fig. 1e) and the quantum-tunnelling rotor
has six-fold rotational symmetry, the complex amplitudes are given
by
\begin{equation}
J_{r}= J\exp {\left[ 3\times \left( i\frac{2\pi}{6} \frac{\Phi}{\phi _0}\right) \right] },
\end{equation}
\begin{equation}
J_{l}= J\exp {\left[ -3\times \left( i\frac{2\pi}{6} \frac{\Phi}{\phi _0}\right) \right] },
\end{equation}
where $J$ is the strength of the tunnelling, $\Phi =SB_{\perp }$ is the magnetic flux that penetrates the quantum-tunnelling rotor, $S$ is the area of the closed loop, $B_{\perp}$ is the magnetic field component perpendicular to the plane of the quantum tunnelling rotor, $\phi _0=h/e$ is a quantum of magnetic flux, $h$ is the Planck constant, and $e$ is the elementary charge.
Using these amplitudes, the wave function of the up orientation is given by
\begin{eqnarray}
\label{eq20}
\psi _{up}(t+\Delta t)&=&\psi _{up}(t)+iJ(e^{i \pi \Phi /\phi _0 }+e^{-i\pi \Phi /\phi _0}) \Delta t\psi _{down}(t)\\
&=&\psi _{up}(t)+i2J\cos (\pi \Phi/\phi _0) \Delta t\psi _{down}(t),
\end{eqnarray}
where the transition probability $P$ from $\psi _{\mathrm{up}}$ to
$\psi _{\mathrm{down}}$ (and vice versa) is calculated as
\begin{equation}
\label{eq2}
P\propto\mid \cos (\pi\Phi /\phi _0)\!\mid ^2=[1+\cos (2\pi\Phi /\phi _0)]/2
\end{equation}
using Fermi's golden rule. The tunnelling dynamics depends on the
magnetic flux with a period given by $\phi _0$.

\subsection*{Detection}
Images were acquired from the $\{1/\sqrt{2},-1/\sqrt{2},0\}$ direction by directing the incident Doppler cooling laser along the $\{1/2,1/2,-1/\sqrt{2}\}$ direction.
The setup and trapped structures are depicted in Figure 4.
A one-dimensional Coulomb chain was observed for strong-confinement conditions (a normal mode frequency of $\omega _x >2\pi\times 1.75$ $\mathrm{MHz}$ in Fig. 1b), 
whereas near intermediate confinement (Figs. 4c and d), the orientation of the triangular structure did not change
during Doppler cooling. The orientations of the triangles in these two cases  could be distinguished within a detection time of 50 ms.

Even under intermediate-confinement conditions, the  triangular orientation was not stable against background-gas collisions (300 K).
However, because of the ultra-high vacuum, the collision rate of such collisions is approximately 1 every few minutes and was thus sufficiently slow that the orientations could be detected.

\subsection*{Stabilisation of the trap radio frequency}
Because the tunnelling frequency depends on the normal mode frequencies, the trap radio frequency (RF) and radial potential must be stabilised.
The amplitude of the trap RF was monitored using a pick-up coil and an RF power meter, and the trap RF was stabilised using a feedback technique.
The fluctuations of the normal mode frequencies were less than a few hundred hertz.

\subsection*{Sideband cooling}
The sideband cooling of a two-dimensional structures in a linear Paul
trap is difficult to achieve because the ions are subject to large
micromotions when they are not located at the node of the trap RF.
However, we were able to cool the rotational mode to near the ground
state using sideband cooling with a red sideband transition laser
($^2\mathrm{S}_{1/2}-{^2\mathrm{D}_{5/2}}$; $729$ nm) and a
quenching laser ($^2\mathrm{D}_{5/2}-{^2\mathrm{P}_{3/2}}$; $854$
nm). The lasers were simultaneously incident on the trap for $5$ ms.
The asymmetry between the red and blue sideband spectra after
sideband cooling (Figure 5) indicated cooling to near
the ground state.

\subsection*{Adiabatic cooling}
The ground state of the rotational mode must be prepared to observe quantum tunnelling, and because the normal mode frequency of the rotational mode is quite small under tunnelling conditions, an extremely low temperature is necessary to achieve this goal.
To obtain a ground-state population larger than 0.1, a temperature of less than $90$ nK is required; this temperature is much lower than the recoil limit in conventional sideband cooling.
Adiabatic cooling, which has been demonstrated with trapped ions\cite{adiabatic}, was therefore performed to achieve the ground state.
Reducing the energy of a single motional quantum $\hbar \omega$ under adiabatic conditions $(\frac{d\omega }{dt}/\omega ^2\ll 1 )$ reduces the temperature, even though the phonon entropy is conserved.
The mean phonon number $\bar{n}$ at temperature $T$ is calculated using boson statistics as follows:
\begin{equation*}
\bar{n}=\frac{1}{\exp{(\hbar \omega /k_B T)}-1},
\end{equation*}
where $\omega$ is the secular frequency of the mode, $k_B$ is the Boltzmann constant, and $\hbar$ is the reduced Planck constant.

In the experiment, the mean phonon number of the rotational mode just after sideband cooling was 0.08, which corresponds to $10$ $\mathrm{\mu K}$.
We reduced the secular frequency adiabatically from $750$ to $0.18$ $\mathrm{kHz}$ to reduce the temperature to $40$ $\mathrm{nK}$, although there was some heating, as indicated by an increase in the entropy of the motional mode.
This temperature was sufficiently cold for quantum tunnelling to be observed.

\subsection*{Heating}
Heating of the trapped ions causes a decrease in the population of the ground state of the quantum-tunnelling rotor. 
For the intermediate-confinement state, we observed a heating rate after sideband cooling that was much less
than 1 quantum/s, and no heating of the rotational mode was observed
within 1 s under these conditions. 
Determining the mean phonon number after a time interval of 1 s was difficult because the other modes were subject to stronger heating. 
This lack of sensitivity can be attributed to the symmetry of the rotational mode.
It is difficult for electric-field noise to heat the rotational mode because the electrodes are separated from the rotor by distances that are much larger than the rotor's own length-scale.
The other modes, e.g., the zigzag mode, could not be cooled by the sideband cooling because of its strong heating.

We measured the phonon number of the rotational mode after ramping down the potential to create
the triangular structure of the quantum-tunnelling rotor and then ramping it back up to its original magnitude.
After the potential was ramped up, the mean phonon number was 8.
This value was nearly independent of the waiting time between ramping down and and up for times of up to 300 $\mathrm{ms}$.
Hence, we speculate that the heating of the rotational mode predominantly occurred during the variation of the potential.
Although the trap frequencies were very low near the critical points, the heating rate for the rotational mode remained small.
This low heating was also likely attributable to the symmetry of the rotational mode.
However, this value depended on the position of the centre of the triangular structure. 
The origins of the heating of the rotational mode may have been trap RF, other hot modes, and various sources of noise.
These heating sources were carefully eliminated by changing the position of the rotor.
Thus, it is possible that fine control of the DC voltages of the linear Paul trap in synchronisation with the variation of the amplitude of the trap
RF may permit the preparation of a quantum-tunnelling rotor that is colder than that of the current experiment.

\subsection*{Magnetic field}
To ensure that the magnetic field was sufficiently stable for sideband cooling,
we used two orthogonal sets of coils to generate the magnetic fields.
The first set created a constant magnetic field ($3.4$ $\mathrm{G}$)
for the sideband cooling and an offset of the magnetic flux, whose
directional vector was $\{1/2,-1/2,1/\sqrt{2}\}$. Here, the normal
vector of the triangular structure was taken to be $\{0,1,0\}$ (see
Figure 4a). The second set of coils created a tunable
magnetic field whose direction was $\{1/2, -1/2, -1/\sqrt{2}\}$ to
change the number of magnetic-flux quanta penetrating the quantum-tunnelling rotor. From the area of the closed loop (37 $\mathrm{\mu
m}^2$) and the direction of the magnetic field, the number of
magnetic-flux quanta that penetrated the quantum rotor was $\Phi /\phi
_0=1.5\pm 0.2$ when there was no current flowing through the tunable coils.
This value is subject to certain errors because of the unknown residual magnetic
fields and the angular error between the coils and the triangular
structure of the ions.

\section*{Fitting}
The fit to the transition-probability data presented in Figure 2b is given by
\begin{equation*}
f(p_0,\nu ,\tau _0, v)=p_0 \times (\frac{1-e^{-(t/T_2)^2}\cos (2\pi\nu t)}{2})+(1-p_0)\times \frac{1-e^{-vt}}{2},
\end{equation*}
where $p_0$ is the population of the rotational ground state of the quantum tunnelling rotor, $\nu$ is the tunnelling frequency, $T_2$ is the coherence time of the tunnelling and $v$ is the average rotational velocity of the rotational excited states.
Fitting parameters of $p_0=0.10\pm 0.02$, $\nu =7.6\pm 0.3$ $\mathrm{Hz}$, $T_2=300\pm200$ $\mathrm{ms}$, and $v =5.4\pm 0.3$ $\mathrm{s}^{-1}$ were used.
The coherence time may have been limited by the heating of the rotational mode and by fluctuations in the tunnelling rate.
According to Figure 2c in the main text, the tunnelling rate $\nu$ was
\begin{equation*}
\nu \sim 21.5 \mathrm{Hz}-0.008 \times (\omega _x-\omega _z)/(2\pi )
\end{equation*}
near the experimental condition $\omega _x-\omega _z\sim 2\pi\times 2$
$\mathrm{kHz}$. If the fluctuation in the radial confinement
is assumed to have been approximately 300 $\mathrm{Hz}$ because of imperfections in the
feedback control of the trap RF, the fluctuations in the tunnelling
frequency were $2.4$ $\mathrm{Hz}$. This indicates that if we consider only tunnelling-rate fluctuations, the coherence
time was $130$, which is consistent with the fitting parameter.

The fit function used to obtain the results presented in Figure 3 in the main text was
\begin{equation}
g(a, \xi ,\theta _0,h)=\frac{a}{2}\cos{(2\pi \xi n+\theta _0)}+h,
\end{equation}
where $n$ is the number of magnetic-flux quanta penetrating the quantum-tunnelling rotor.
The fitting parameters were $a=0.06\pm 0.02$, $\xi =0.99\pm 0.07$, $\theta _0=0.28\pi \pm 0.22 \pi$, and $h =0.15\pm 0.01$.

% {\bf{Supplementary Information}} is linked to the online version of the paper at www.nature.com/nature
\section*{Acknowledgements}

The authors thank Shmuel Nussinov, Yasunobu Nakamura and Takuya
Higuchi for their useful discussions. This work was supported by the MEXT
KAKENHI ``Quantum Cybernetics" Project, a Grant-in-Aid for Young
Scientists (Grant No. 25800181), and the Japan Society for the
Promotion of Science (JSPS) through the ``Funding Program for
World-Leading Innovative R\&D on Science and Technology (FIRST
Program)," which was initiated by the Council for Science and Technology
Policy (CSTP). A.N. was supported in part by the Japan Society for
the Promotion of Science.

\section*{Author Contributions}
A.N. performed the experiment with assistance from K.T. and S.U.,
Y.S. provided theoretical guidance regarding the AB-effect
experiment, and S.U. headed the project. All of the authors contributed to
writing the manuscript.

{\bf{Author Information}} Reprints and permissions information are
available at www.nature.com/reprints. The authors declare no
competing financial interests. Correspondence and requests for
materials should be addressed to A.N.
(noguchi@qe.ee.es.osaka-u.ac.jp).

\pagebreak

{\bf Figure 1 \textbar\! The quantum tunnelling rotor.} {\bf a,} CCD
images of the two degenerate triangular crystal structures. {\bf b,}
Calculated energies of the four lowest eigenmodes. The solid curve
corresponds to the rotational mode, and the dashed curve corresponds to the
zigzag mode. The dot-dashed and dotted curves represent the energies of
the centre-of-mass modes of the radial and axial directions,
respectively. 
The left- and right-hand insets display the sets of displacement vectors for the rotational and zigzag modes, respectively.
The normal-mode frequency in the axial direction is
$\omega _z=2\pi\times 1.119$ $\mathrm{MHz}$. {\bf c,} The quantum-tunnelling
rotor, indicated the potential-energy and pattern diagrams of the
quantum tunnelling of a 3-ion triangular structure.
{\bf d,} The solid blue curve represents the potential energy as a function of the rotational angle.
The red (purple dashed) triangle corresponds to one of the bound states in the well, as denoted by `up' (`down') in {\bf a}, and the dashed red (purple) curve represents the corresponding wave function.
{\bf e,} A charged quantum rotor in a magnetic field.
The sign of the AB phases is dependent on the direction of rotation, where $\theta (=6\pi (\Phi /\phi _0))$ is the AB phase for tunnelling.

\text{ }

{\bf Figure 2 \textbar\! Quantum tunnelling in the rotor.} {\bf a,}
Experimental time sequences. M1: measurement of the initial state,
SB: sideband cooling of the rotational mode, and M2: measurement of the final state. {\bf b,} The correlation between the initial and final
states from 400 experiments per data point as a function of the waiting time $\tau$. The fit function is the sum
of a sinusoidal curve and an exponential decay. The error bars represent the standard deviations of binomial distributions.
{\bf c,} The tunnelling rate as a function of the radial confinement. The solid curve represents a numerical calculation performed using the experimental parameter $\omega _z =2 \pi\times 1.119$ $\mathrm{MHz}$. 
The vertical error bars represent the standard errors of fitting.
The horizontal error bars correspond to the standard deviation of the amplitude of the trap RF.

{\bf Figure 3 \textbar\! The Aharonov-Bohm effect during quantum tunnelling.} 
The transition probability as a function of the magnetic flux produced by the tunable coils expressed in magnetic quantum units($\phi _0 = 4.14\times 10^{-15} \mathrm{Wb}$).
We performed 800 experiments per data point.
The error bars are statistical errors calculated by assuming binomial distributions of the results. The solid curve represents a cosine fit to the results.

{\bf Figure 4 \textbar\! Experimental system.} {\bf
a}, The relative geometry of the ion trap and the imaging
system. Images were acquired of three $^{40}\mathrm{Ca}^+$ ions under various conditions: {\bf b},
under strong-confinement conditions ($\omega _x  = 2\pi\times 2.1 \mathrm{MHz}$), {\bf c,} under slightly stronger than intermediate confinement ($\omega _x=2\pi\times 1.65\mathrm{MHz}$), {\bf d,} under intermediate confinement conditions ($\omega _x  = 2\pi\times 1.523 \mathrm{MHz}$), and {\bf e,} near the critical point($\omega _x  = 2\pi\times 1.119 \mathrm{MHz}$).

{\bf Figure 5 \textbar\! Sideband spectra of the rotational mode.} 
Rotational blue and red sideband spectra {\bf a,} before and {\bf b,}
after sideband cooling. The number of excitations of the three ions
is presented as a function of the frequency. 
We performed 200 experiments per data point.
The error bars represent the standard deviations of binomial distributions.

\text{ }
\pagebreak

%% Figure 1%%%%
\begin{figure}[h]
\includegraphics[width=15cm,angle=0]{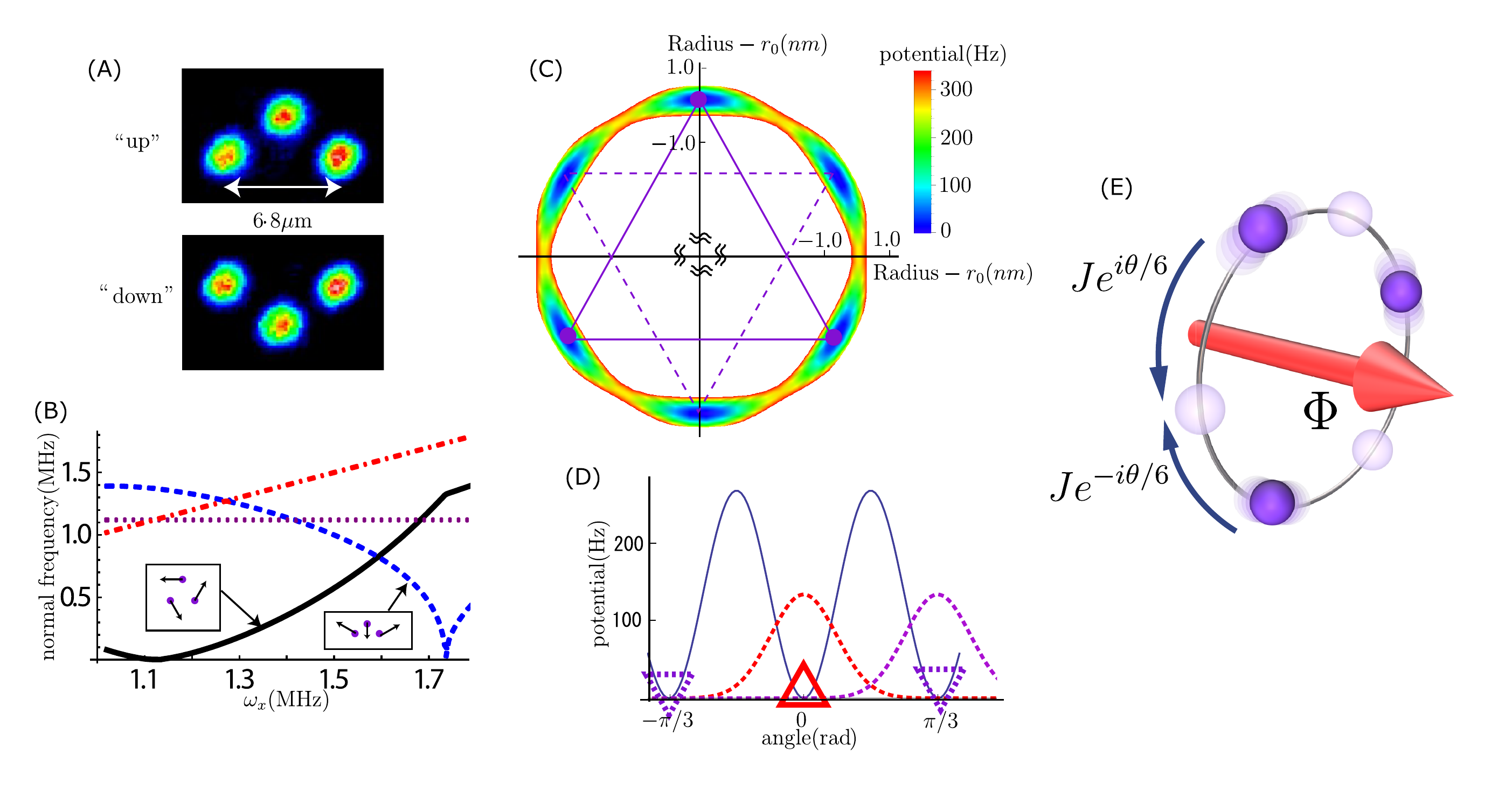}
\label{fig1}
\caption{}
\end{figure}
%%%%%%%%%%%

\pagebreak

%% Figure 2%%%%
\begin{figure}[h]
\begin{center}
\includegraphics[width=20cm,angle=0]{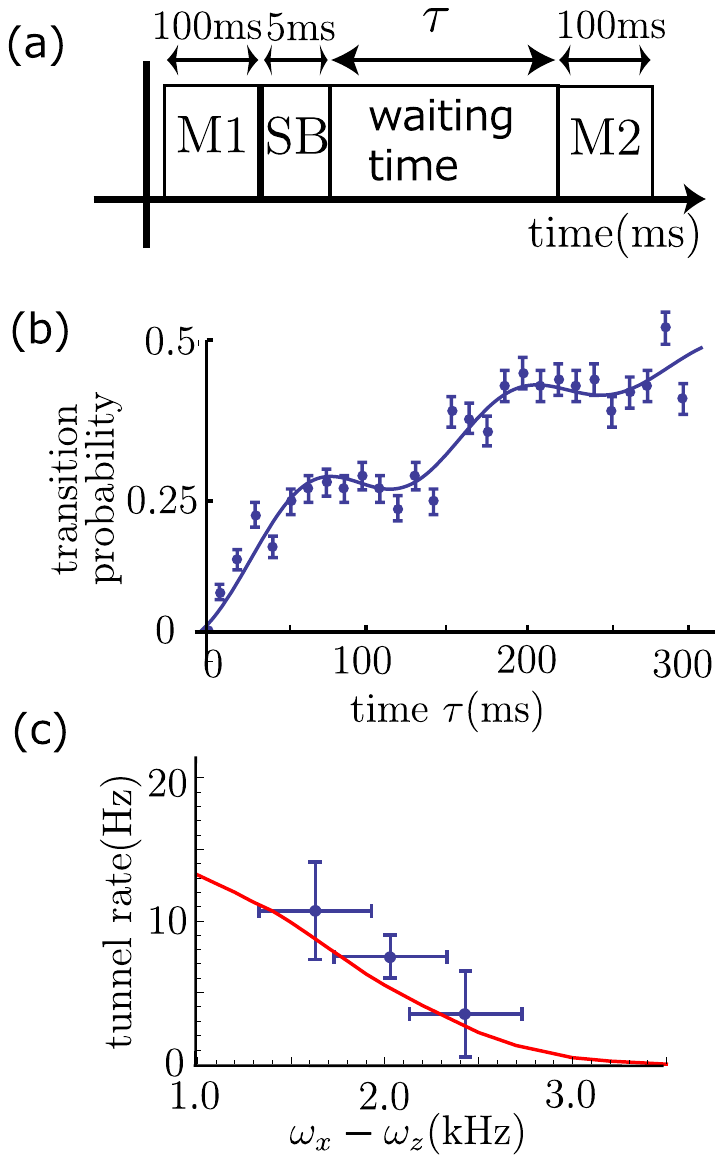}
\label{fig2}
\caption{}
\end{center}
\end{figure}
%%%%%%%%%%%

\pagebreak

%% Figure 3%%%%
\begin{figure}[h]
\begin{center}
\includegraphics[width=20cm,angle=0]{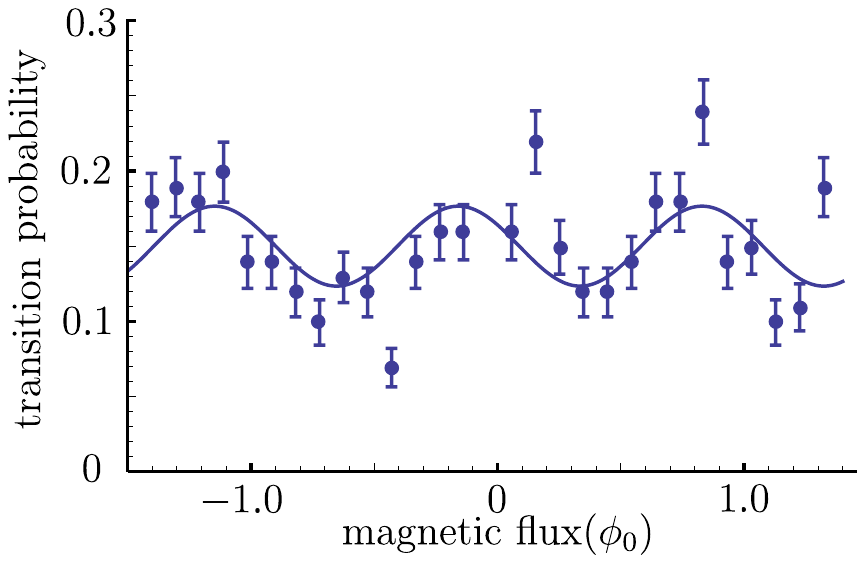}
\label{fig3}
\caption{}
\end{center}
\end{figure}
%%%%%%%%%%%

\pagebreak

%% Figure B%%%%
\begin{figure}[h]
\begin{center}
\includegraphics[width=18cm,angle=0]{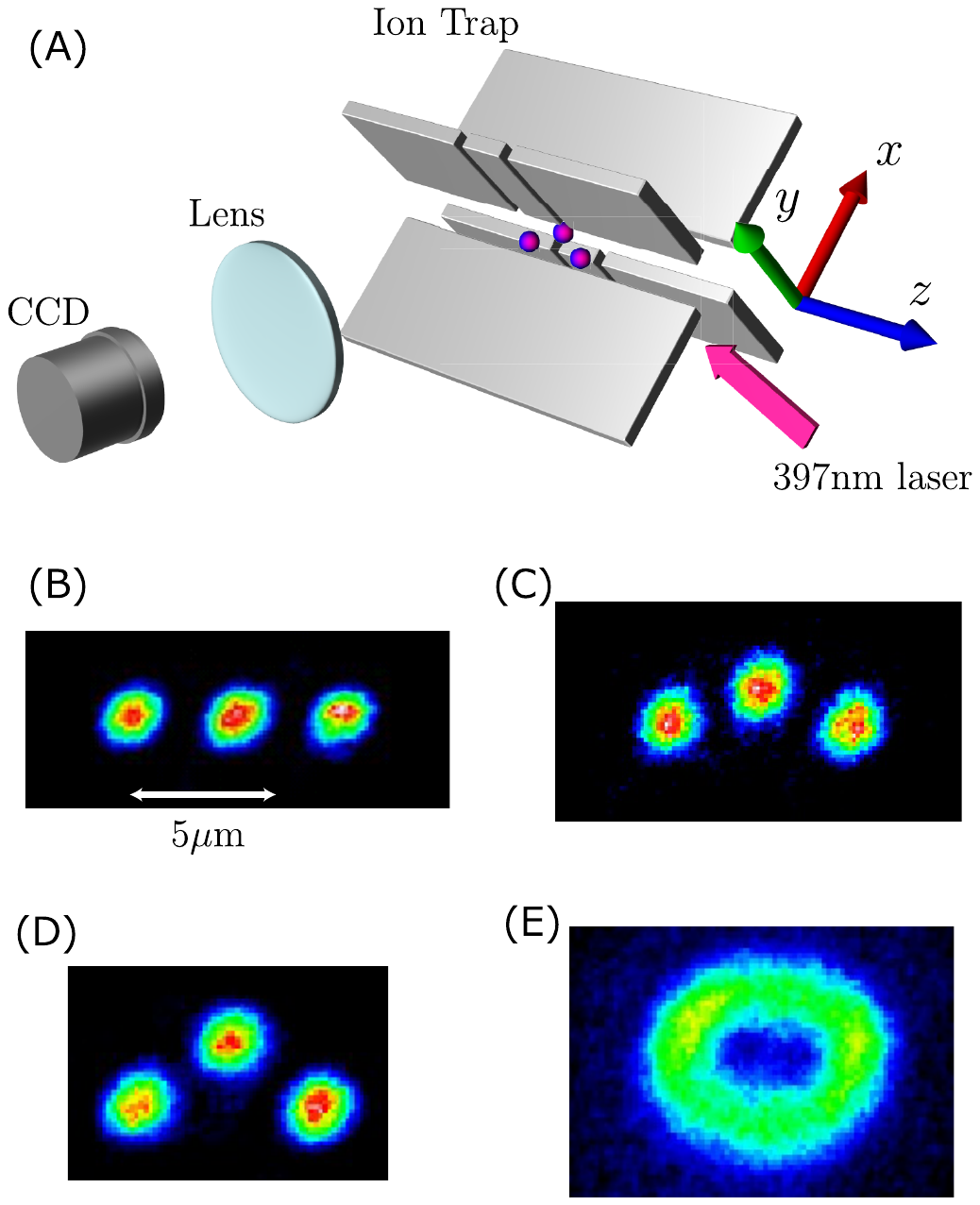}
\label{figB}
\caption{}
\end{center}
\end{figure}
%%%%%%%%%%%

\pagebreak

%% Figure C%%%%
\begin{figure}[h]
\begin{center}
\includegraphics[width=19cm,angle=0]{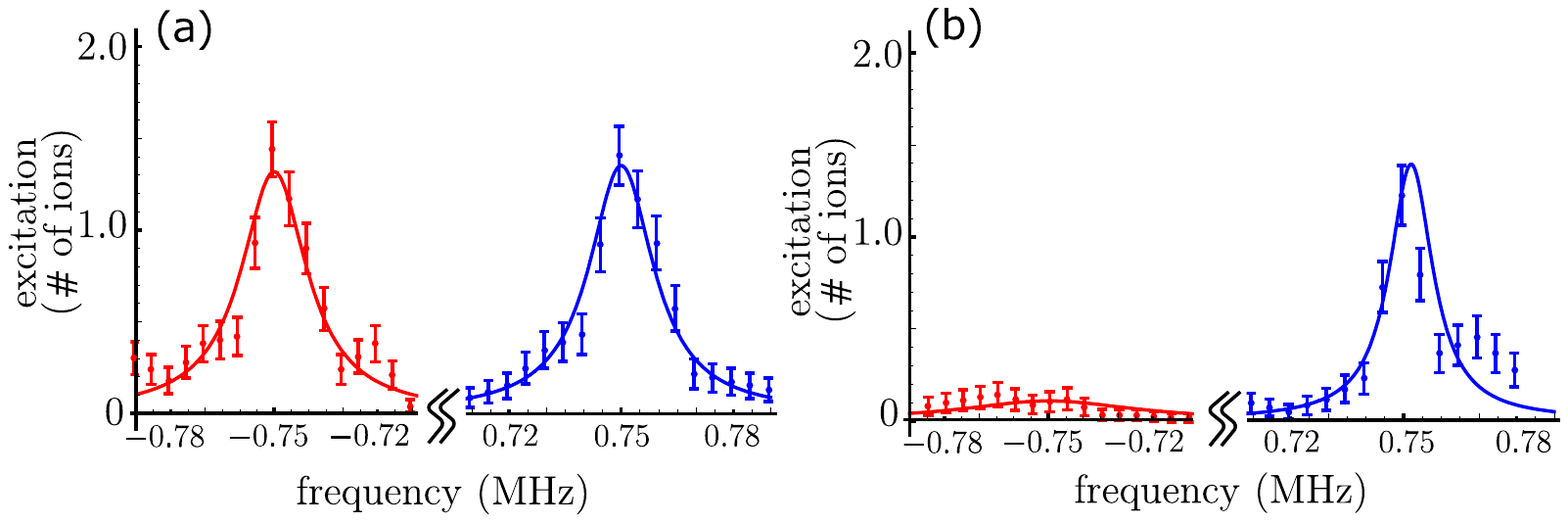}
\label{figC}
\caption{}
\end{center}
\end{figure}
%%%%%%%%%%%

\end{document}